\documentclass[aip,amsmath,amssymb,reprint,floatfix,pof]{revtex4-1}
\usepackage{graphicx}
\usepackage{dcolumn}
\usepackage{bm}
\usepackage[utf8]{inputenc}
\usepackage[T1]{fontenc}
\usepackage{mathptmx}
\usepackage{etoolbox}
\usepackage{verbatim}
\usepackage{xcolor}
\usepackage{soul}

\usepackage{hyperref}
\hypersetup{colorlinks,
 linkcolor=blue,%
 citecolor=blue,%
 urlcolor=blue
}

\newcommand{\mbr}{\mathbf{r}}
\newcommand{\mbx}{\mathbf{x}}
\newcommand{\mbk}{\mathbf{k}}
\newcommand{\mbu}{\mathbf{u}}

\begin{document}

\title{Energy spectra and fluxes of turbulent rotating Bose-Einstein condensates in two dimensions}

\author{Anirudh Sivakumar 
}
\affiliation{Department of Physics, Bharathidasan University, Tiruchirappalli 620 024, Tamil Nadu, India}

\author{Pankaj Kumar Mishra}
\affiliation{Department of Physics, Indian Institute of Technology Guwahati, Guwahati 781039, Assam, India}

\author{Ahmad A. Hujeirat}
\affiliation{Interdisciplinary Center for Scientific Computing, The University of Heidelberg, 69120 Heidelberg, Germany}

\author{Paulsamy Muruganandam 
}
\affiliation{Department of Physics, Bharathidasan University, Tiruchirappalli 620 024, Tamil Nadu, India}

\date{\today}

\begin{abstract}
We investigate the scaling of the energy cascade in a harmonically trapped, turbulent, rotating Bose-Einstein condensate (BEC) in two dimensions. We achieve turbulence by injecting a localized perturbation into the condensate and gradually increasing its rotation frequency from an initial value to a maximum. The main characteristics of the resulting turbulent state depend on the initial conditions, rotation frequency, and ramp-up time. We analyze the energy and the fluxes of kinetic energy by considering initial profiles without vortices and with vortex lattices.
In the case without initial vortices, we find the presence of Kolmogorov-like scaling ($k^{-5/3}$) of the incompressible kinetic energy in the inertial range.  However, with initial vortex lattices, the energy spectrum follows Vinen scaling ($k^{-1}$) at transient iterations. For cases with high rotating frequencies, Kolmogorov-like scaling emerges at longer durations. We observe positive kinetic energy fluxes with both initial states across all final frequencies, indicating a forward cascade of incompressible and compressible kinetic energy.
\end{abstract}

\maketitle

\section{Introduction} 
\label{sec:intro}

Superfluid turbulence, also known as quantum turbulence (QT), has been a topic of interest within the scientific community since the time of Feynman~\cite{Feynman1955}. Feynman proposed that one could visualize QT as a chaotic tangle of quantized vortex filaments. Building upon this, Vinen made an experimental observation that turbulence in superfluid ($^4\mathrm{He}$), in general, can be sustained by the mutual friction between quantized vortices and the normal fluid component of the superfluid, driven by a constant temperature gradient in the horizontal direction \cite{Vinen1957}.

Vortices play a central role in understanding fluid turbulence, and a comparative study between classical turbulence (CT) and QT could provide a better understanding of the nature of vortices formed in these systems. Specifically, in the case of 3D turbulence, both CT and QT exhibit similar macroscopic and statistical properties. They both display a similar nature of the energy cascades in the large length scale, particularly in the inertial range, as confirmed numerically in earlier studies~\cite{Nore1997, Parker2005, Maurer1998}. However, understanding the genesis of turbulence and further energy cascade from the first principle in CT is relatively rigorous due to the continuous and chaotic nature of the vortices. In contrast, vortices observed in QT possess vortex cores with definite sizes and circulations, making them an ideal prototype for investigating turbulence dynamics \cite{Barenghi2008, Barenghi2014, Khomenko2015, Streltsov2015}.

Despite experimental studies on QT through superfluid helium, which show significant promise in understanding the phenomena \cite{Hall1956}, manipulating and studying these quantized vortices in such superfluids remains a challenging experimental task. In the last few decades, Bose-Einstein condensates (BECs), superfluids existing below a critical temperature and free of viscous effects, have emerged as a novel platform for exploring the role of quantized vortices in generating the QT. BECs offer advantages, such as compressibility, weak atomic interactions, the ability to fine-tune atomic parameters, and the availability of new experimental methods for probing and studying superfluid flow \cite{Madeira2020}.

Several studies in the past reveal chaotic dynamics of quantum vortices aided by the complex nature of the vortex reconnection responsible for the turbulence in QT \cite{Donnelly1991, Schwarz1985}. For instance, Aref investigated the influence of vortex number on the transition of a quantum fluid from chaotic to turbulent flow \cite{Aref1983}. Cornell and co-workers have pioneered developing the cooling technique that accelerates the rotation of an ultracold $^{87}$Rb gas and nucleates vorticity in a Bose-Einstein condensate~\cite{Haljan2001}. People have also achieved the same effect by mechanically rotating an anharmonically confined condensate \cite{Madison2000, AboShaeer2001, Hodby2001}. However, Parker and Adams were the first to numerically demonstrate the occurrence of turbulence, with direct energy scaling, upon attainment of the crystallization of a vortex lattice in 2D BECs \cite{Parker2005}. Using the laser beam stirring technique, Neely et al. demonstrated the first experimental confirmation of quantum turbulence in 2D BECs \cite{Neely2013} through the observation of the formation and evolution of a disordered distribution of vortices in highly oblate annular BECs. Other methods for generating quantum turbulence in atomic BECs include introducing a stirring potential in the condensate and evolving the turbulent state into a vortex lattice through vortex-sound interactions \cite{Parker2005},  phase-imprinting a disordered vortex lattice onto a 2D BECs \cite{Madarassy2009} and combined-axis rotation around two axes for the 3D trapped BECs \cite{Kobayashi2007}. 

A feature of 2D CT is the presence of energy cascades, in which kinetic energy flows from larger to smaller length scales (direct energy cascade) or from smaller to larger length scales (inverse energy cascade) in an inertial range. The corresponding incompressible kinetic energy spectrum follows Kolmogorov scaling in Fourier space, i.e. $\varepsilon_{\mathrm{kin}}^i(k) \sim k^{-5/3}$. In QT, both inverse and direct cascades of the energy spectrum have been reported depending on the initial configuration. This scaling has been observed in 3D QT numerically by Kobayashi et al. \cite{Kobayashi2005, Kobayashi2005a} and has also been observed in 3D CT. Further similarities between QT and CT have been confirmed numerically by solving the Gross-Pitaevskii equation \cite{Nore1997, Parker2005}.

In a turbulent BECs, the energy cascade very much depends on the means through which turbulence is being generated in the system. As an example, the turbulent BECs in 2D generated using the rotating paddles show the forward cascade of the energy with Kolmogorov-like scaling~\cite{Das2022}. Numasato et al. demonstrated Kolmogorov-like scaling for BECs undergoing decaying homogeneous 2D quantum turbulence generated by random phase initial conditions \cite{Numasato2010}. Their study also revealed how vortices and sound waves introduced into the BECs can significantly influence the observed spectra. However, the turbulence generated through the stationary grid with obstacles in the 2D BECs exhibits the inverse cascade of the energy with Kolmogorov-like scaling \cite{Reeves2013}. Similar features have been observed with decaying and unforced turbulence \cite{Mueller2020}. There are some works that show the presence of Vinen-like scalings for the kinetic energy spectra ($\varepsilon_{\mathrm{kin}}^i\sim k^{-1}$) for the fast rotating 2D BECs~\cite{AmetteEstrada2022, Estrada2022}. Similar scalings have been reported for decaying 2D BECs turbulence in which the cluster of vortices breaks into multiple ones \cite{Cidrim2017, Marino2021}. 

Previous studies on QT have primarily employed unstable initial condensates, which tended to produce self-sustaining or decaying turbulence phenomena during real-time evolution. Furthermore, earlier investigations into 2D and 3D QT have typically focused on Gross-Pitaevskii (GP) equations without a rotational term, with rotational forcing achieved through anharmonic and time-varying potentials. For instance, Estrada and colleagues recently carried out a study on 3D condensates in the rotational frame \cite{AmetteEstrada2022}. Given the substantial disparities between dimensional regimes for QT in non-rotating GPEs, our objective is to explore these differences by simulating rotational QT in 2D.

In our present work, we consider a stable initial condensate prepared via imaginary-time iterations. We then generate the turbulence using a time-dependent rotational frequency in a perturbed central barrier. We choose two initial states, viz., the vortex-free and the vortex lattice. In the first case, we identify that the incompressible kinetic energy exhibits Kolmogorov-like scaling with a forward energy cascade. The Kolmogorov-like scaling becomes more pronounced at higher rotational frequency. However, in the latter case, we find the Vinen-like scaling for the energy spectrum with a forward cascade at a lower rotational frequency. For both cases, the negative particle number flux suggests the transfer of the particle from small to large length scale, which becomes more significant at high rotational frequency.

The organization of this paper is as follows: In Sec. \ref{sec:model}, we present the mathematical model for simulating rotating BECs with the details of the perturbed central barrier used to generate turbulence, which also includes the protocols used to introduce the angular frequency and detailed forms of the relevant energy, such as incompressible and compressible energy spectra. We use these spectra to characterize the energy cascade in the later section. Sec. \ref{sec:num} deals with numerical simulation results that we obtained by solving the mathematical model mentioned in Sec. \ref{sec:model}. In Sec. \ref{sec:vlesscase}, we discuss the energy spectra and fluxes of the turbulent state, which we attain with an initial state having no vortex present in the fluid, while in Sec. \ref{sec:vlatticecase}, we present the results for the situation when the vortices pinned on a lattice are used as an initial condition to get the turbulent state. Finally, we conclude our investigation in Sec. \ref{sec:summary}.


\section{Description of the model and energy spectra}
\label{sec:model}

We consider a quasi-two-dimensional condensate confined strongly in a transverse direction rotating with an angular frequency $\Omega$. The dynamical equation of the condensate in the non-dimensional form is given by
\begin{align}
\mathrm{i}\frac{\partial\psi}{\partial t} = & \left[-\frac{1}{2}\nabla^2 + V(\mbr) + g_{2D} \lvert \psi \rvert^2 - \Omega (t) L_z \right] \psi,
\label{eq:gpe}
\end{align}
where $\psi \equiv \psi(\mathbf{r}, t)$ denotes the condensate wave function, with $\mathbf{r} \equiv (x, y)$, $\nabla^2$ is the two-dimensional Laplace operator defined as $\nabla^2 \equiv \partial_x^2 + \partial_y^2$, and $V(\mathbf{r})$ is the external potential, which includes the harmonic trap along with the central circular barrier. The nonlinear term $g_{2D} = 4\pi aN/\sqrt{2\pi}d_z$ represents the interaction strength between the atoms, where $N$ is the total number of atoms, $a$ is the s-wave scattering length, and $d_z$ corresponds to the axial width of the trap. Here, $\Omega(t)$ represents the rotational frequency, and $L_z = \mathrm{i}\hbar(y \partial_x - x \partial_y)$ is the $z$ component of angular momentum. 

In Eq.~(\ref{eq:gpe}), the unit length is measured in terms of the harmonic oscillator length $l = \sqrt{\hbar/(m \omega)}$, time is measured in units of $\omega^{-1}$, and trap frequency is measured in terms of $\omega$.
Here, we consider two types of initial configurations: (i) vortex-less and (ii) vortex lattice. For the former case while we consider $N=1\times10^4$ and $g_{2D} = 100$, for the later $N=2\times10^4$ and $g_{2D} = 200$ have been taken. With the choice of the above parameters, we obtain the scattering length $a \approx 3.8 a_0$ for both cases. The BECs should have a strong axial confinement along $z$ axis so as to restrict the dynamics within the $x-y$ plane by implementing trap frequencies $\omega_x \sim 2\pi \times 33 \mathrm{Hz}$, $\omega_y \sim 2\pi \times 33 \mathrm{Hz}$, $\omega_z \sim 2\pi \times 1.5 \mathrm{MHz}$. This configuration provides a trap frequency of $\omega \sim 2\pi\times 116.4 Hz$, harmonic oscillator length $l \approx 1\mu$m and unit time $\omega^{-1} \approx 1.3$ms. The oscillator length along the $z$ direction would be $l_z \approx 0.28 \mu$m. Experimentally, for a condensate of $^{87}\mathrm{Rb}$ atoms, the desired scattering length can be accessible by tuning the magnetic fields utilizing the Feshbach resonance~\cite{Marcelis2004, Bauer2009}. 

To induce turbulence in the rotating BECs, it requires some sort of perturbation that generates the disordered structure of the vortex lattice from the initial Abrikosov vortex lattice~\cite{Mueller2020, Estrada2022, AmetteEstrada2022}. In our work, we exploit a perturbed barrier to induce turbulence. We evolve the condensate using a gradually increasing rotational frequency via a perturbed barrier. The external potential (in dimensionless form) is given by
\begin{align}
V(\mbr) = \frac{1}{2}\left(\gamma^2x^2 + \nu^2y^2\right) + V_{\mathrm{B}}(\mbr), \label{eq:trap}
\end{align}
where $\gamma=1$ and $\nu=1$ are the aspect ratios of the harmonic trap along the radial direction. The perturbed barrier ($V_B$) is given by
\begin{align}
V_{\mathrm{B}}(\mbr) = 
\begin{cases}
V_0, & r < R + A\left[ \sin \left(\alpha \Theta \right) + \sin \left(\beta \Theta + \delta \right)\right] \\
0, & \mathrm{otherwise},
\end{cases}
\label{eq:pot}
\end{align}
with $A$, $R$, $V_0$ are the perturbation amplitude, barrier radius, and barrier height, respectively, $r = \lvert \mbr \rvert = \sqrt{x^2 + y^2}$ and $\Theta  = \arctan(y/x)$. For our simulations, we set the amplitude to be $A = 1.25 \, l$, the barrier radius as $R = 0.5\, l$, and we choose the barrier height as $V_0 = 20 \hbar \omega$ and $\alpha = 10$. We consider two configurations of perturbations: (i) periodic perturbation for which $\beta = 20$ and $\delta = 0.4$ and (ii) quasiperiodic perturbation for which $\beta = 5 \times \left(1 + \sqrt{5} \right)$ and $\delta = 0.4$ are chosen. Note that the perturbed potential considered in the paper can be realized in the laboratory experiment by considering the superposition of two optical lattices with the frequencies $\alpha$ and $\beta$ in the angular coordinate direction.
\begin{figure}[!ht]
\centering
\includegraphics[width=0.99\linewidth]{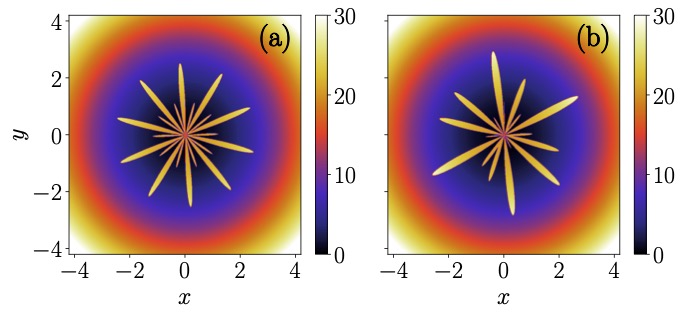}
\caption{Plots showing central portion of the harmonic trap superimposed with the perturbed barrier: (a) periodic perturbation with $\alpha = 10$, $\beta = 20$ and $\delta = 0.4$ (b) quasiperiodic perturbation with $\alpha = 10$, $\beta = 5 \times \left(1+\sqrt{5} \right)$ and $\delta = 0.4$.}
\label{fig:barrier}
\end{figure}
In Figs.~\ref{fig:barrier}(a) and (b) we depict the central portion of the harmonic trap \eqref{eq:trap} superimposed with the perturbed barrier \eqref{eq:pot} for periodic and quasiperiodic perturbations, respectively.

We carry out the simulation by gradually increasing the rotation frequency from the initial value to the final one. We used the following form of the temporal rotational frequency $\Omega(t)$:
\begin{align}
\Omega(t) = 
\begin{cases}
\displaystyle \Omega_{0} + \left( \Omega_{\mathrm{f}} - \Omega_{0}\right) \sin^2 \left( \frac{\pi t}{2 T_r} \right), & \mathrm{if}\; 0 \le t \le T_r, \\
 \Omega_{\mathrm{f}}, & \mathrm{if} \; t > T_r, \\
\end{cases} 
\label{eq:omegaprof}
\end{align}
\begin{figure}[!ht]
    \centering
    \includegraphics[width=0.99\linewidth]{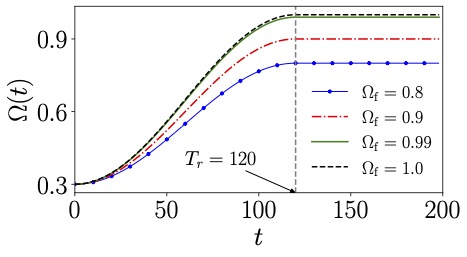}
    \caption{A typical representation of the variation of rotating frequency $\Omega(t)$ with respect to time as implemented for vortex less initial initial state for different $\Omega_{\mathrm{f}}$. Here initial frequency is fixed to $\Omega_0=0.3$.}
    \label{fig:rotfreq}
\end{figure}
where $\Omega_0$ and $\Omega_{\mathrm{f}}$ represent the initial and final angular frequencies, respectively, and $T_r$ represents the time interval during which the rotation frequency is smoothly increased from an initial value of $\Omega_0$ to a final value of $\Omega_f$. In the simulation we have fixed $T_r=120$ and considered the different values of $\Omega_f$. Figure~\ref{fig:rotfreq} illustrates a typical representation of the temporal increment of angular frequency $\Omega$ for different final frequencies ($\Omega_f$) with the same initial frequency $\Omega_0=0.3$.

\subsection{Spectra and Flux Calculations}

The flow of incompressible kinetic energy across wavenumbers and its relationship to vortex dynamics are central and crucial to understanding the nature of the QT at different lengths and time scales. To obtain the scaling laws for energy spectra, we decompose the condensate kinetic energy into compressible and incompressible parts and analyze the distribution of kinetic energy due to vortex lines and sound over length scales, as shown by Nore et al.~\cite{Nore1997}.

To obtain these spectra, we perform a Madelung transformation on the wavefunction and obtain a density weighted velocity field as
\begin{align}
\mbu (\mbr) = \sqrt{n}(\mbr)\mathbf{v}(\mbr), \label{eq:wvf}
\end{align}
where $n(\mbr) = \left\vert\psi(\mbr)\right\vert^2$ is the particle density and $\mathbf{v}(\mbr) = \hbar/m \nabla\theta(\mbr)$ is the superfluid velocity and $\theta(\mbr)$ is the corresponding phase.

We decompose this weighted velocity field into the hydrodynamic and quantum pressure components. This definition of the density-weighted velocity field allows for a Helmholtz decomposition into the hydrodynamics components, $\mbu(\mbr) = \mbu(\mbr)^i + \mbu(\mbr)^c$. The incompressible and compressible components satisfy %
\begin{subequations}
\begin{align}
\nabla\cdot\mbu^i(\mbr) & = 0, \label{eq:vincomp} \\
\nabla\times\mbu^c(\mbr) & = 0, \label{eq:vcomp}
\end{align} 
\end{subequations}
respectively. This decomposition is performed by applying a Fourier transformation to the velocity fields. Apart from these hydrodynamic components, the other relevant quantity quantum pressure velocity field is defined as
\begin{align}
\mbu^q(\mbr) = \frac{\hbar}{m}\nabla\sqrt{n(\mbr)}. \label{eq:qp}
\end{align}
Quantum pressure becomes significant when the condensate density varies sharply, such as near the vortex cores. For 2D condensates, it provides a more useful quantitative measure for the vortex number.
The quantum pressure velocity field does not contribute to the physical velocity field but does possess the dimensions of velocity and satisfies the condition $\nabla \times \mbu^q = 0$. Next, we express the total kinetic energy as
\begin{align}
 E_{\mathrm{kin}} = E_{\mathrm{kin}}^i + E_{\mathrm{kin}}^c + E_{\mathrm{kin}}^q ,
 \label{eq:tke}
\end{align}
where $E_{\mathrm{kin}}^i$, $E_{\mathrm{kin}}^c$ and $E_{\mathrm{kin}}^q$ represent, respectively, incompressible, compressible, and quantum pressure contributions of the kinetic energy.
Following Parseval's theorem, different energy components can be represented as
\begin{align}
E^{\zeta}_{\mathrm{kin}} = \frac{m}{2}\int d^2\mbk\lvert\tilde{\mbu}^{\zeta}(\mbk)\rvert^2,
\label{eq:parseval}
\end{align}
where $\zeta \in \{i,c,q\}$ being incompressible, compressible and quantum pressure components of the kinetic energy, respectively and 
\begin{align}
\tilde{\mbu}^{\zeta}(\mbk) = \frac{1}{2\pi}\int d^2\mbr e ^{-i\mbk\cdot\mbr}\mbu^{\zeta}(\mbr)
\label{eq:ftransform}
\end{align}
Transforming Eq.~(\ref{eq:parseval}) into the cylindrical coordinates in Fourier space ($k$ space) for a 2D condensate, we obtain
\begin{align}
E_{\mathrm{kin}}^{\zeta} = \int_0^{\infty} dk \varepsilon_{\mathrm{kin}}^{\zeta}(k) .
\label{eq:ekinzeta}
\end{align}
Usually, the spectra $\varepsilon_{\mathrm{kin}}^{\zeta}$ are computed by binning the data in $k$ space and further summing over an angular interval, an approximation that holds good in small-$k$ regimes. We use the analytical evaluation of $k$-space integrals and its numerical implementation developed by Bradley and colleagues \cite{Bradley2022}. The method involves an angle-averaged Wiener-Khinchin theorem relating the spectral densities to an associated correlation function. So the spectra $\varepsilon_{\mathrm{kin}}^{\zeta}$ from Eq.~(\ref{eq:ekinzeta}) can be written as~\cite{Bradley2022}
\begin{align}
\varepsilon_{\mathrm{kin}}^{\zeta}(k) = \frac{m}{2}\int d^2\mbx\Lambda_2(k,\lvert\mbx\rvert)C[\mbu^{\zeta}, \mbu^{\zeta}](\mbx),
\label{eq:specden}
\end{align}
where $\Lambda_2(k,\lvert \mbx \rvert) = (1/2\pi) k J_0(k\lvert\mbx\rvert)$ is the 2D kernel function, involving the Bessel function $J_0$ and $C[\mbu^{\zeta}, \mbu^{\zeta}](\mbx)$ represents the two-point auto-correlation function in position for a given velocity field. The above relation implies that for any of the position-space fields $\mbu^{\zeta}$, there exists a spectral density [see Eq.~(\ref{eq:specden})], i.e., equivalent to an angle-averaged two-point correlation in $k$ space.
Next, we consider the relevant quantities to characterize the energy cascade, such as the kinetic energy and particle number fluxes of the turbulent state, which provide a quantitative idea about the flow of the energy and the particle from one scale to another. One can use these spectral densities to compute the incompressible kinetic energy and particle fluxes. The incompressible kinetic energy and density flux equations are given by \cite{GarciaOrozco2020}
\begin{subequations}
\begin{align}
\Phi_{\varepsilon_{\mathrm{kin}}^i}(k) & = -\frac{d}{dt}\int_{k_0}^{k} \varepsilon_{\mathrm{kin}}^i(k^{'}) dk^{'}, \label{eq:kinflux} \\ 
\Phi_{{n}}(k) & = -\frac{d}{dt}\int_{k_0}^{k} n(k^{'}) dk^{'}, \label{eq:denflux}
\end{align}
\end{subequations}
where $n(k)$ and $\varepsilon_{\mathrm{kin}}^i(k)$ correspond to the density and incompressible kinetic energy spectra, respectively, and $k_0\sim2\pi/L$ represents the largest length scale, with $L$ being the length of the box. The positive nature of the flux indicates the transfer of energy from the large to small scale, while the negative sign suggests the transfer from the small to large scale of the turbulent state. 

\section{Numerical simulations}
\label{sec:num}

We numerically solve the Gross-Pitaevskii (GP) equation (\ref{eq:gpe}) using the split-step Crank-Nicolson method \cite{Hujeirat2005, Hujeirat2008, Muruganandam2009, Vudragovic2012, YoungS.2017, Kumar2019}. We use imaginary time propagation to generate the ground states with and without vortices and real-time propagation to investigate the condensate dynamics. In all simulations presented in the paper, we use a two-dimensional grid with $512 \times 512$ points, where the space step is $dx = dy = 0.05$ and the time step is $dt = 0.001$. This choice of grid and step sizes ensures numerical convergence and the desired accuracy in our simulations. We have generated the ground state for both the vortex-free and vortex-lattice cases using imaginary time propagation and then evolved them further in real time for a finite angular frequency.

In this work, we intend to analyze the energy spectra and fluxes of the turbulent state of the rotating condensate using various initial setups. In the following sections, we present a systematic analysis of the effect of rotational frequency on the turbulence attained through different initial configurations of the vortices. We emphasize the role of the initial state in achieving the various turbulent states in QT.

\subsection{Energy transfer and fluxes for Vortex-Less initial state}
\label{sec:vlesscase}
We begin our analysis by considering the initial state without vortices and demonstrate the appearance of quantum turbulence by increasing the rotation beyond a threshold frequency.  %
\begin{figure*}[!ht]
\centering
\includegraphics[width=0.95\linewidth]{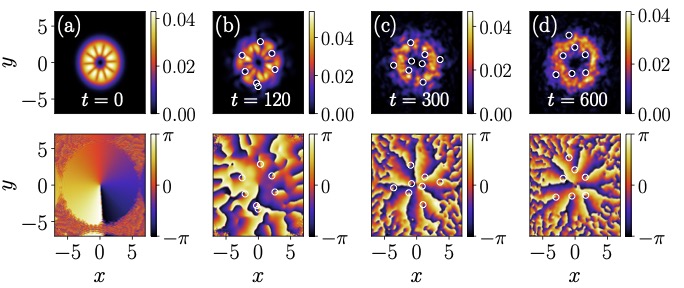}
\caption{Pseudo color representation of the condensate density at the different instants of time with vortex-less initial condition undergoing turbulence for the cases when the final rotating frequency $\Omega_{\mathrm{f}} = 1.0$ was introduced at $t=120$. (a) Density at $t = 0$, (b) Density at $t=120$, (c) Density at $t=300$, (d) Density at $t=600$. The bottom panels depict the corresponding phases. White circles guide the eye to identify the vortex positions. Most of the vortices appear to cluster around the central perturbation.}
\label{fig:densities}
\end{figure*}
To accomplish this, we generate an initial condensate profile of the rotating BECs using a perturbed barrier by evolving the condensate with a constant rotating initial frequency of $\Omega_{0}$ in imaginary time. We select the value of $\Omega_0$ based on the critical rotating frequency, which determines the angular frequency above which vortices start appearing in the condensate. As we consider $N=1 \times 10^4$ atoms and nonlinearity strength $g_{2D} = 100$, the critical frequency comes as $\Omega_{c} \approx 0.4$ also considered in the Ref.~\cite{Kumar2019}. Following this, we have chosen $\Omega_0 = 0.3 < \Omega_c$ to prepare a vortex-less initial condensate ground state using the imaginary time propagation. Subsequently, we employ a real-time evolution approach to delve into the dynamics of the condensate for various values of the final angular frequency, $\Omega_{\mathrm{f}}$. Furthermore, we utilize the evolved state to compute the incompressible kinetic energy ($\varepsilon_{\mathrm{kin}}^i(k)$), compressible kinetic energy ($\varepsilon_{\mathrm{kin}}^c(k)$), and density ($n(k)$) spectra in  Fourier space with $k$ as a wavenumber. We also quantify the the cascade energy using the fluxes of the turbulent condensate at various rotational frequencies using the methods described in the previous section [see Eqs.~\eqref{eq:kinflux} and \eqref{eq:denflux}].

As the system attains the turbulent state, the fluctuations appear at several length and time scales. For our spectrum analysis we identify three characteristic length scales, namely, Thomas-Fermi radius $R_{TF}=\sqrt{2\mu}$, intervortex separation $\ell_0 = 1/\sqrt{n_v}$, and condensate healing length $\xi = 1/\sqrt{\mu}$, where $\mu$ is the chemical potential and $n_v$ is the number of vortices per unit area~\cite{AmetteEstrada2022}.  %
\begin{table}[!ht]
\caption{Estimate of different characteristic length scales $R_{TF}$, $\xi$, and $\ell_0$ of turbulent condensates computed for various final rotation frequencies with the vortex-less initial state. All the reported values are averaged over time interval $t=120$ to $t=600$. }
\label{tab:vless_scales}
\centering 
\begin{tabular}{l|r|r|r}
\hline
\hline
\multicolumn{1}{c|}{$\Omega_{\mathrm{f}}$} & \multicolumn{1}{c|}{$R_{TF}$} & \multicolumn{1}{c|}{$\xi$}  & \multicolumn{1}{c}{$\ell_0$} \\
\hline
\hline
$0.85$ & $4.140$ & $0.342$ & $1.498$\\
$0.9$  & $4.313$ & $0.328$ & $1.529$\\
$0.99$ & $4.579$ & $0.310$ & $1.534$\\
$1.0$  & $4.645$ & $0.305$ & $1.584$\\
\hline
\end{tabular}
\end{table}
In Table \ref{tab:vless_scales} we provide the estimated values of  $R_{TF}$, $\xi$, and $\ell_0$ for different final rotating frequencies, averaged over the time interval $t=120-600$ when the turbulence state is attained with the vortex-less initial state. We find that all the scales show increasing trend with increase in the final rotation.

To show the evolution of the vortex-less initial state to turbulent state in Fig.~\ref{fig:densities}, we depict the snapshots of the condensate density profiles at various time instants as the final rotational frequency $\Omega_{\mathrm{f}}$ as achieved at a time span of $t=120$. Figure~\ref{fig:densities}(a) shows a vortex-less initial condensate and the shape of the central barrier with a significant perturbation. As the rotation frequency attains a value $\Omega_{\mathrm{f}}$ at  $t=120$ vortices begin to appear in the condensate, though they are not yet arranged in a vortex lattice formation as shown in Fig.~\ref{fig:densities}(b). Subsequently, as time progresses, the vortices start converging towards the central barrier and organize into an ordered lattice, as illustrated in Figs.~\ref{fig:densities}(c)-(d). The density profiles clearly show a slow development of compressible turbulence accompanied by consistent vortex generation beyond $t = 120$.

\begin{figure}[!ht]
\centering
\includegraphics[width=\linewidth]{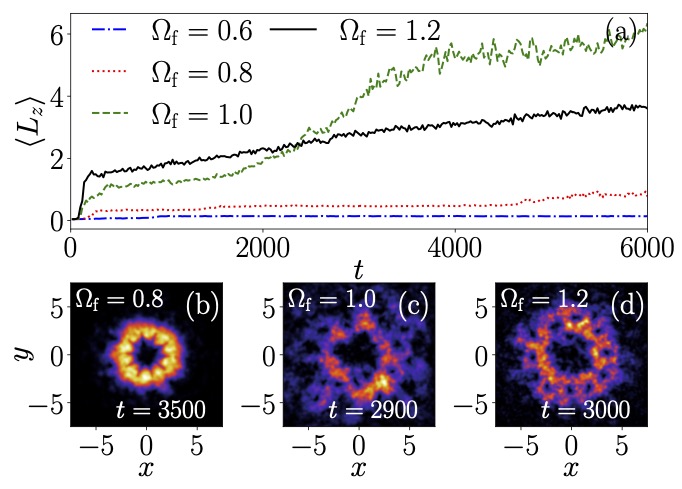}
\caption{(a) Variation of the mean angular momentum ($\langle L_z\rangle$) with respect to time for $\Omega_{\mathrm{f}} = 0.6$ (blue-dash dotted line), $\Omega_{\mathrm{f}} = 0.8$ (red-dotted line), $\Omega_{\mathrm{f}} = 1.0$ (green-dashed line), $\Omega_{\mathrm{f}} = 1.2$ (black-solid line), evaluated for a condensate with a vortex-less initial condition. The condensate attains turbulence for $\Omega_{\mathrm{f}} \geq 0.8$. (c), (d) and (e) show the condensate density profiles at $t=3500$ [red-dotted line in (a)], $t=2900$ [green-dashed line in (a)], and $t=3000$ [black-solid line in (a)], respectively.}
\label{fig:amless}
\end{figure}

To understand different transient states of the turbulence, we show the temporal evolution of the mean angular momentum ($\langle L_z \rangle$) for different final angular frequencies ($\Omega_{\mathrm{f}}=0.6$, $0.8$, $1$, and $1.2$) in Fig.~\ref{fig:amless}. The steady-state value of $\langle L_z \rangle$ very much depends upon the $\Omega_{\mathrm{f}}$; it increases with an increase in $\Omega_{\mathrm{f}}$, indicating the increase in the number of vortices and having more turbulent states at higher $\Omega_{\mathrm{f}}$. For $\Omega_{\mathrm{f}} < 0.8$, there is no change in the angular momentum with time, indicating no generation of vortices and thereby providing a lower bound for $\Omega_{\mathrm{f}} \approx 0.8$ for our simulations below which we do not have any vortex in the condensate. %
For $0.8 \lesssim \Omega_{\mathrm{f}} \lesssim 1$, we find that the angular momentum settles to a finite value at shorter times  but continues to increase at longer duration due to persistent rotation. Notably, when the rotation frequency is $\Omega_{\mathrm{f}}=1$, we observe a stable $\langle L_z \rangle$ value within the time window of  $t \sim 400-1500$. However, beyond this window, the value of $\langle L_z \rangle$ increases and surpasses the $\langle L_z\rangle$ value associated with $\Omega_{\mathrm{f}}=1.2$ (around $t \sim 2500$). This increase in the angular momentum may be attributed to the generation of a large number of vortices, resulting from the appearance of degenerate Landau levels when the rotating frequency resonates with the trap frequency~\cite{Fetter2009}.
Once $\langle L_z \rangle$ saturates the corresponding real-time density snapshot of the condensate density shows the clustering of the vortices around the central barrier as shown in the bottom panel of Fig.~\ref{fig:amless} for different angular frequencies ($\Omega_{\mathrm{f}}=0.8$, $1$ and $1.2$). 
For final rotation frequencies greater than the trap frequency ($\Omega_{\mathrm{f}}>1$), the angular momentum does not exhibit a tendency to reach a steady value even at a shorter duration and also does not show consistent scaling behaviour.

After characterizing the different turbulent states at various angular frequencies, we then focus on analyzing the nature of the kinetic energy spectra and their corresponding fluxes at different angular frequencies during the time interval where the condensate exhibits the presence of vortices clustered around the central barrier. We compute the temporal averages of the spectra and fluxes for the different components of kinetic energy from the instance when the final rotational frequency ($\Omega_{\mathrm{f}}$) is achieved at $t = T_r = 120$ until the end of the simulation at $t = 600$. The kinetic energy spectra provide insights into the characteristic structure and arrangement of vortices and their overall effect in inducing turbulent fluctuations in the condensate.
\begin{figure}[!ht]
\centering
\includegraphics[width=\linewidth]{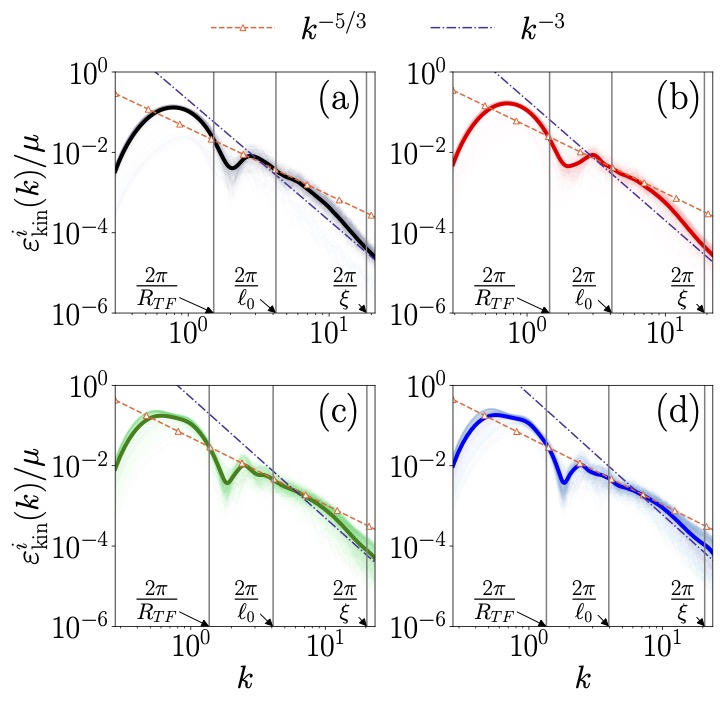}
\caption{Time-averaged spectra of incompressible kinetic energy for the vortex-less case, in the time range $t=120$ to $t = 600$ (a)-(d) For $\Omega_{\mathrm{f}} =0.85$, $\Omega_{\mathrm{f}} = 0.95$, $\Omega_{\mathrm{f}} = 0.99$, $\Omega_{\mathrm{f}} = 1.0$ respectively, it exhibits $k^{-5/3}$ accompanied by $k^{-3}$ scaling at larger $k$ values.}
\label{fig:ikinspectra}
\end{figure}

\begin{figure}
\centering
\includegraphics[width=\linewidth]{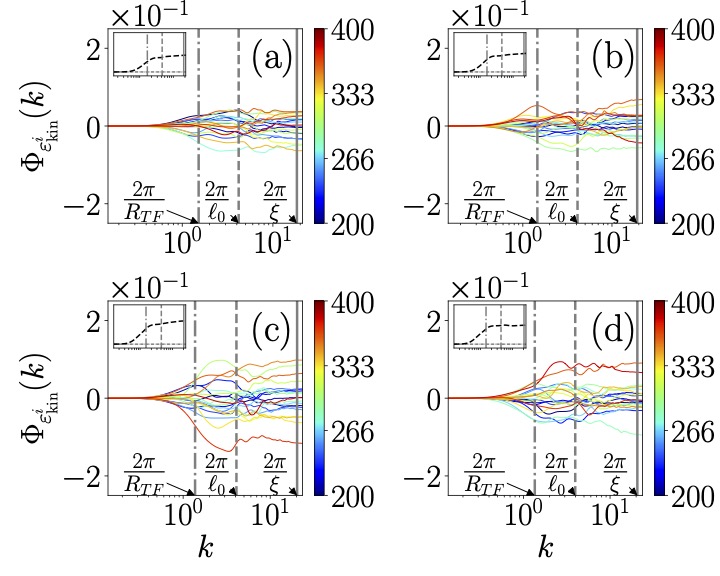}
\caption{Plots of the incompressible kinetic energy fluxes for the vortex-less case at the different instants, as given in the color bar. Fluxes are computed for the time range $t = 200$ to $t = 400 $ at final rotation frequencies: (a) $\Omega_{\mathrm{f}} = 0.85$, (b) $\Omega_{\mathrm{f}} = 0.9$, (c) $\Omega_{\mathrm{f}} = 0.95$, and (d) $\Omega_{\mathrm{f}} = 1.0$. The various scales are $k_{R_{TF}}=  2 \pi / R_{TF}$ (gray-dash dotted line), $k_{\ell_0} = 2\pi/\ell_0$ (gray-dashed line) and $k_{\xi} = 2\pi/\xi$ (gray-solid line). The insets display time-averaged incompressible kinetic energy flux for each rotational frequency (magnitude $\sim 10^{-2}$) for the same $k$ range as the main plot}. 
\label{fig:kinflux}
\end{figure}

\begin{figure}[!ht]
\centering
\includegraphics[width=\linewidth]{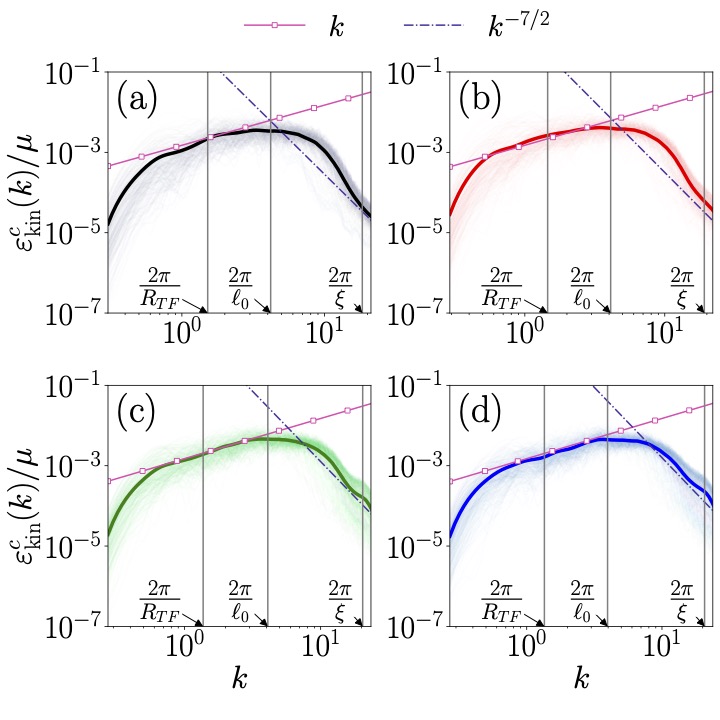}
\caption{Time-averaged spectra of compressible kinetic energy for the vortex-less case, in the time range $t=120$ to $t = 600$ for the condensate rotating at: (a) $\Omega_{\mathrm{f}} = 0.85$, (b) $\Omega_{\mathrm{f}} =0.9$, (c) $\Omega_{\mathrm{f}} = 0.99$, (d) $\Omega_{\mathrm{f}} = 1.0$. All cases display $k$ scaling, whose range extends for higher rotation frequencies, indicating strong thermalization of the condensate.} 
\label{fig:ckinspectra}
\end{figure}

For our spectral analysis, we define the healing length scale in $k$ space $k_{\xi} = 2\pi/\xi$, along with the Thomas Fermi radius $k_{R_{TF}}=2\pi/R_{TF}$ and inter-vortex length $k_{\ell_0}=2\pi/\ell_0$. The characteristics scales for different $\Omega_f$ are given in the Table~\ref{tab:vless_scales}. 
Figure.~\ref{fig:ikinspectra} illustrates the scaled incompressible kinetic energy spectra for different rotational frequencies $\Omega_{\mathrm{f}} = 0.85, 0.95, 0.99$ and $1.0$ in the wavenumber space. For all frequencies $\Omega_{\mathrm{f}}$, the spectrum at large wave number ($k\sim 2\pi/\xi$) appears to fall with a scaling $k^{-3}$ a typical behaviour for the energy spectrum due to the presence of the vortex core~\cite{Bradley2012}. However,  the energy spectrum falls as $k^{-5/3}$ on both sides of the inter-vortex distance scale $k\sim 2\pi/\ell_0$ as shown in Fig.~\ref{fig:ikinspectra}(a). Upon increase in the rotation frequency, the scaling appears to fit better in the given region, indicating the chaotic spatial distribution of the vortices [cf. Figs.~\ref{fig:ikinspectra}(b)-(d)]. The scaling behaviour appearing at scales larger than the inter-vortex distance indicates the turbulence arising due to the collective motion of vortices, and the cascade continues in the $2\pi/\ell_0 < k < 2\pi/\xi$ region as well, where individual vortex dynamics dominate. The observation of Kolmogorov-like scaling for the incompressible kinetic energy is generally associated with the faster decay of the vortex tangles in 3D superfluid turbulence and equivalently transfer of energy towards the large scale during the decay of the vortex tangle~\cite{Walmsley2008, Baggaley2012}. However, establishing a similar kind of feature for the 2D QT is quite a subtle task. As we look at the energy spectra for $\Omega_{\mathrm{f}} > 1.0$, we find a continuous generation of vortices prohibits the system from achieving the steady state. Rotating at such elevated frequencies leads to the disruption of confinement in the condensate, facilitating its continuous expansion. This expanded state fosters the creation of an environment conducive to the formation of additional vortices. Without consistent turbulence dynamics and the continual breakdown of vortices, this scenario imposes a constraint on the final rotational frequency that can be achieved. In addition to spectral analysis of the incompressible kinetic energy, we also examine the incompressible density profiles of the kinetic energies in real space, as given in the Appendix \ref{app1}. It suggests the localization of the incompressible kinetic energy at the outer periphery of the condensate, which becomes more prominent at a higher angular frequency as shown in Fig.~\ref{fig:ienerden}.

\begin{figure}[!ht]
\centering
\includegraphics[width=\linewidth]{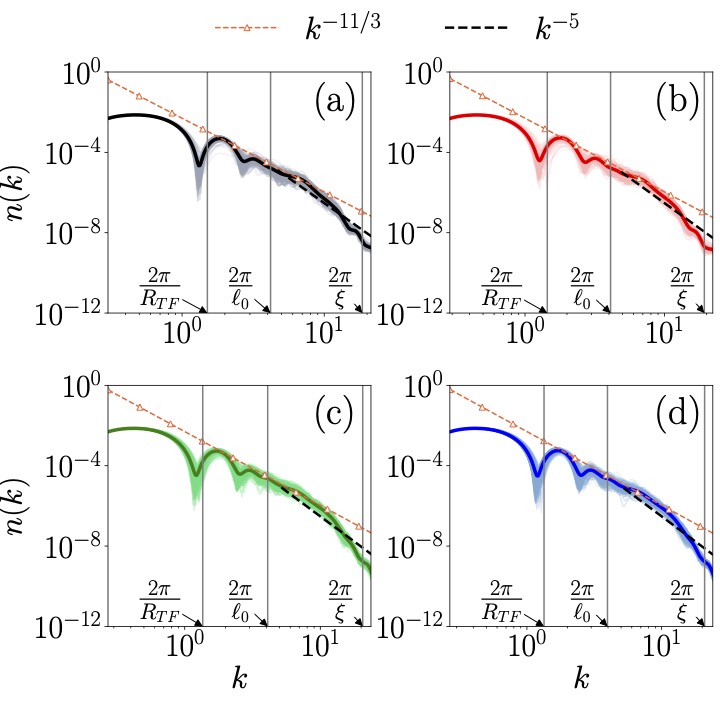}
\caption{Plots of the average density spectra for the time range $t = 120$ to $t = 600$ with final rotation frequencies (a) $\Omega_{\mathrm{f}} = 0.85$, (b) $\Omega_{\mathrm{f}} = 0.9$, (c) $\Omega_{\mathrm{f}} = 0.99$, (d) $\Omega_{\mathrm{f}} = 1.0$. All figures show $k^{-11/3}$ scaling, indicating the onset of the Kolmogorov cascade.}
\label{fig:denspectra}
\end{figure}

In Fig.~\ref{fig:kinflux}, we show the flux corresponding to the incompressible component of the kinetic energy calculated using the expression defined in Eqs.~(\ref{eq:kinflux}). Figures~\ref{fig:kinflux}(a)-(d) denote the flux for the rotational frequencies $\Omega_{\mathrm{f}}=0.85, 0.9, 0.95$, and $1$, respectively, at different instances where the pseudo colour bars indicate the time. We find that for all the frequency ranges, the flux fluctuates between positive and negative with a breathing frequency of $2.0\omega$. For $k$ values $k >2\pi/\ell_0$, the amplitude of the flux does not change with the wave number for a given instant of time, complementing the nature of cascade due to the presence of the Kolmogorov scalings in this range.
However, as we analyze the variation of flux averaged over a time interval ($200 < t < 400$) with wave number, we find that it appears to remain positive for all the final rotation frequencies, as shown with the dashed black line. This feature indicates the forward cascade of the incompressible part of the kinetic energy in the IR as well as the UV range consistent with the earlier studies related to the energy cascade in 2D quantum turbulence~\cite{Numasato2010}. The role of rotation here is simply to increase the rate of energy transfer from one scale to another.

In Fig.~\ref{fig:ckinspectra}, we show the time-averaged scaled compressible kinetic energy in the wave number space for different rotation frequencies ($\Omega_{\mathrm{f}}=0.85$, $0.9$, $0.99$, and $1$). For all the frequencies, we find that the compressible spectrum behaves as $k$ for small wave numbers, while at large wave numbers, it exhibits $k^{-7/2}$ scaling. Note that in general, the compressible kinetic energy attains the thermal equilibrium scaling, i.e.,  $\varepsilon_{\mathrm{kin}}^i(k) \sim k$ for the decaying turbulence where all the energy transferred to the compressible part of the energy~\cite{Numasato2010}. However, in the rotating turbulence case, we find that the incompressible part of the kinetic energy dominates over the other components, and thus there is the generation of the vortices. This is the reason that on the scale below the healing length ($\xi$), we find the presence of the $\varepsilon_{\mathrm{kin}}^i(k)\sim k^{-7/2}$. As we analyze the distribution of the compressible density of the kinetic energies in real space, we find that initially localized compressible kinetic energy near the central barrier evolves and get distributed evenly in the condensates as shown in the Fig.~\ref{fig:cenerden}.

\begin{figure}
\centering
\includegraphics[width=\linewidth]{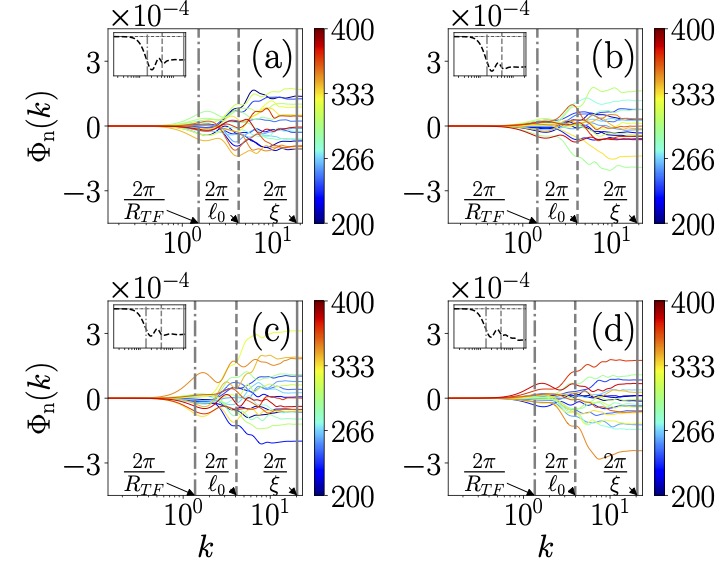}
\caption{Density flux profiles for the vortex-less case. The plots are taken are averaged over the time range $t = 200$ to $t = 400 $ at final rotation frequencies: (a) $\Omega_{\mathrm{f}} = 0.85$, (b) $\Omega_{\mathrm{f}} = 0.9$, (c) $\Omega_{\mathrm{f}} = 0.99$, and (d) $\Omega_{\mathrm{f}} = 1.0$. The inset displays the time average of the density flux (magnitude $\sim 10^{-6}$) for the corresponding rotational frequency for the same $k$ range as the main plot.
}
\label{fig:denflux}
\end{figure}

In Fig.~\ref{fig:denspectra}, we show time-averaged density spectra for the different final rotation frequencies ($\Omega_{\mathrm{f}}=0.85,0.9, 0.99$ and $1.0$). As the incompressible component of the kinetic dominates over the compressible counterpart, we can have $k^2 n(k) \sim \varepsilon^i_{\mathrm{kin}}(k)$, which yields $k^{-11/3}$ and $k^{-5}$ scalings for the density in the $k$ values  $ k  < 2\pi/\ell_0$ and UV range ($k\sim 2\pi/\xi$). For all the frequencies, we find that the particle density spectra follow these scalings at their respective ranges depicted in Fig.~\ref{fig:denspectra}.

To understand the detailed nature of the transfer of the particle from one scale to another, in Fig.~\ref{fig:denflux}, we show the particle density flux ($\Phi_\mathrm{n}(k)$) for different instants of time where we use color codes to illustrate the flux variation with wave number at that instant. Like incompressible kinetic energy flux, density flux also oscillates between positive and negative with time for a given range of the wavenumber. The oscillation frequency is the same as that for the kinetic energy flux. The density flux exhibits an increasing behaviour in the IR range while it attains a constant value in the UV range. As we analyze the time-averaged density flux variation in the spectrum, we find that it takes on negative values for all rotational frequencies, as shown in the average dashed black lines of the respective plots, indicating an inverse cascade of particles from small to large scales. We can attribute these particular features to the aided features of the rotation, which is quite evident from the increasing magnitude of the time-averaged flux upon an increase in the rotational frequency.  %
\subsection{Energy transfer and fluxes for Vortex-lattice initial state}
\label{sec:vlatticecase}
After presenting the detailed nature of the spectrum and fluxes for the vortex-less initial condition, we consider another type of initial state with the presence of vortex lattices. This initial condensate has the same barrier parameters as those considered for the vortex-less case. To introduce vortices into the system, we first prepare an initial condensate in imaginary time with an appropriate choice of $\Omega_0$ that yields the vortices arranged on the lattice. 
\begin{figure*}[!ht]
\centering
\includegraphics[width=0.95\linewidth]{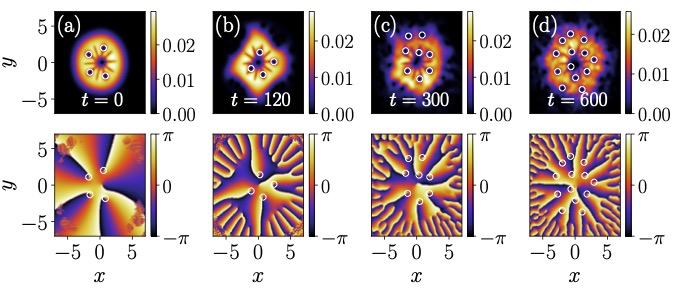}
\caption{Snapshots of the condensate density at the different instants of time: (a) $t = 0$, (b) $t=120$, (c) $t=300$, (d) $t=600$, with the vortex-lattice initial state as condensate starts rotating with  $\Omega_{\mathrm{fin}}=0.85$ at $t=120$. The other parameters are the same as those in Fig.~\ref{fig:densities}.}
\label{fig:latdensities}
\end{figure*}
\begin{figure}[!ht]
\centering
\includegraphics[width=\linewidth]{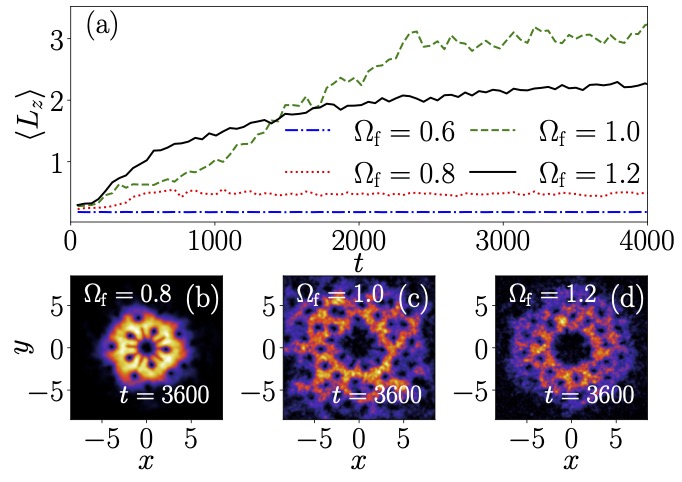}
\caption{Variation of the expectation value of angular momentum with respect to time for different $\Omega_{\mathrm{f}}$ values, evaluated for a condensate with vortex-lattice initial condition. The condensate undergoes turbulence for $\Omega_{\mathrm{f}} \geq 0.8$. The insets show condensate density profiles for $t=3600$ for $\Omega_{\mathrm{f}}=0.8$ (red dotted line), $\Omega_{\mathrm{f}}=1.0$ (green dashed line), and $\Omega_{\mathrm{f}}=1.2$ (black solid line).}
 \label{fig:amlattice}
\end{figure}
An important difference from the vortex-less state here is the atom number and interaction strengths, which we certainly need to increase for the present case in comparison to those for the vortex-less case. Here, we consider $N = 2 \times 10^4$ and $g_{2D} = 200$, with a critical frequency of $\Omega_{c}\approx 0.3$, above which vortices are generated and arranged on the lattice in the condensate. Here, we have chosen a larger nonlinearity strength ($g_{2D}$) than those for the vortex-less initial condition to accommodate the vortex lattice in the condensate. To illustrate the dynamics with a vortex-lattice initial profile clearly and consistently, we consider the rotation frequency as $\Omega_0 = 0.6$  for the results presented below. We have also presented the time averaged values of the Thomas-Fermi radius $R_{TF}$, healing length $\xi$ and intervortex distance $\ell_0$ for various final rotation frequencies, $\Omega_{\mathrm{f}}$, in Table.~\ref{tab:vlattice_scales}. %
\begin{table}[!ht]
\caption{Estimate of $R_{TF}$, $\xi$, and $\ell_0$ values computed for various final rotation frequencies for the vortex-lattice case, averaged over the time interval $t=280$ to $t=380$ for $\Omega_{\mathrm{f}}=0.8$ and $t=120$ to $t=220$ for $\Omega_{\mathrm{f}} = 0.825,0.85,0.89$.}
\label{tab:vlattice_scales}
\centering
\begin{tabular}{l|r|r|r}
\hline
\hline
\multicolumn{1}{c|}{$\Omega_{\mathrm{f}}$}  & \multicolumn{1}{c|}{$R_{TF}$} & \multicolumn{1}{c|}{$\xi$}  & \multicolumn{1}{c}{$\ell_0$} \\
\hline
\hline
$0.8$   & $3.968$ & $0.356$ & $2.487$\\
$0.825$ & $3.990$ & $0.354$ & $2.500$\\
$0.85$  & $4.009$ & $0.353$ & $2.369$\\
$0.9$   & $4.028$ & $0.351$ & $2.380$\\
\hline
\end{tabular}
\end{table}
\begin{figure}[!ht]
\centering
\includegraphics[width=\linewidth]{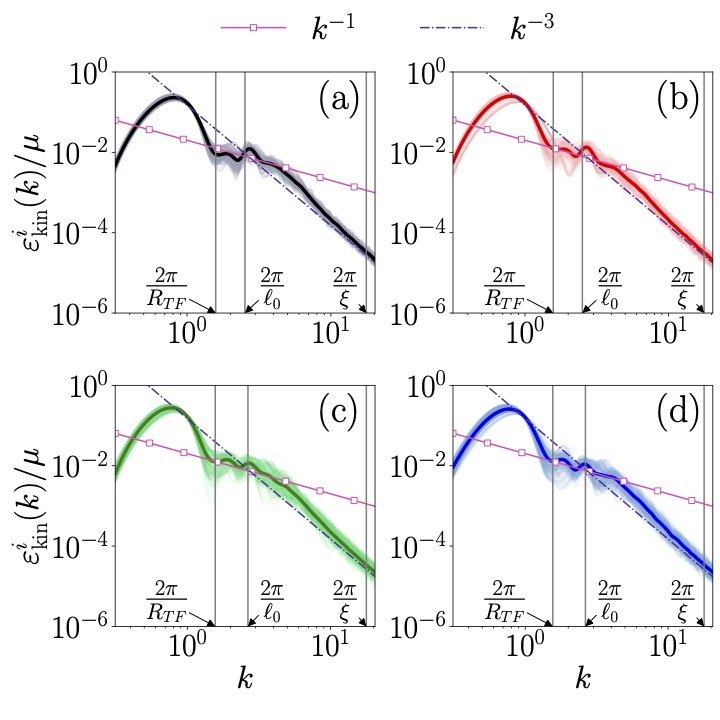}
\caption{Time-averaged incompressible kinetic energy spectra for vortex-lattice initial condition: (a) $t = 280$ to $t = 380$ at $\Omega_{\mathrm{f}} = 0.8$,  from $t = 120$ to $t = 220$ for (b) $\Omega_{\mathrm{f}} = 0.825$, (c) $\Omega_{\mathrm{f}} = 0.85 $, (d) $\Omega_{\mathrm{f}}=0.9$. All the plots display $k^{-1}$ scaling in an inertial range, along with an enstrophy cascade $k^{-3}$.}
\label{fig:latikinspectra}
\end{figure}

\begin{figure}[!ht]
\centering%
\includegraphics[width=\linewidth]{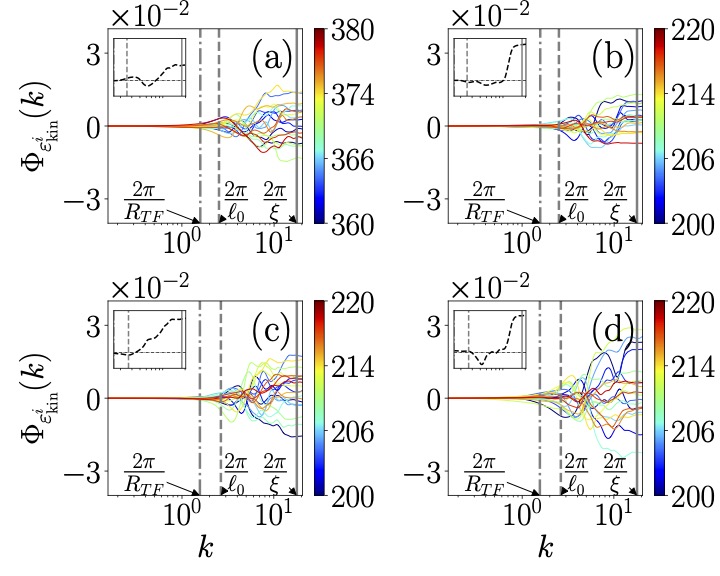}
\caption{Incompressible kinetic energy flux for the vortex-lattice case for (a) $t=360$ to $t=380$ at $\Omega_{\mathrm{f}} = 0.8$, for $t=200$ to $t=220$ at (b) $\Omega_{\mathrm{f}} = 0.825$, (c) $\Omega_{\mathrm{f}} = 0.85$, and (d) at $\Omega_{\mathrm{f}} = 0.9$. The inset displays the time average of the kinetic energy flux (magnitude $\sim 10^{-3}$) for each rotational frequency plotted in the range $2\pi/R_{TF} < k < 2.3\pi/\xi$.}
\label{fig:latkinflux}
\end{figure}

\begin{figure}[!ht]
\centering
\includegraphics[width=\linewidth]{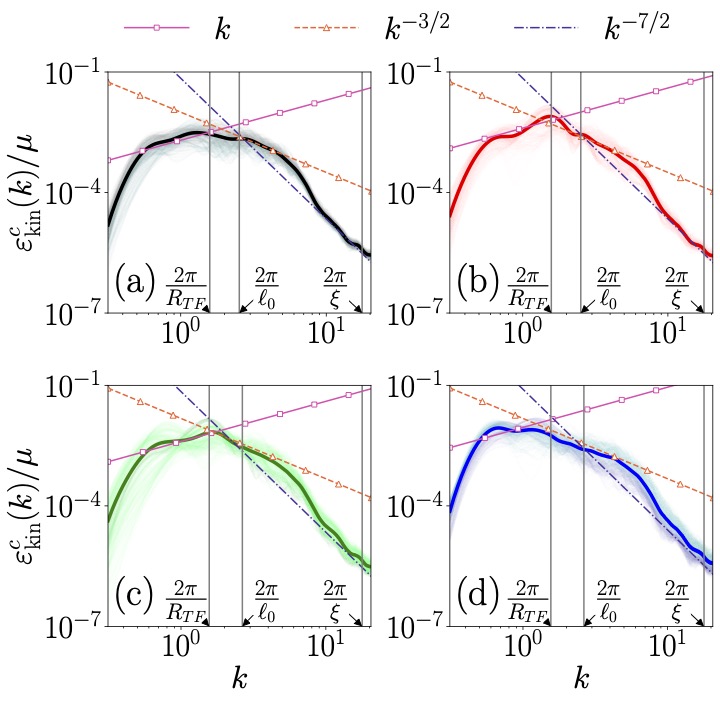}
\caption{Time averaged compressible kinetic energy spectra: (a) $t = 280$ to $t = 380$ at $\Omega_{\mathrm{f}}=0.8$, for $t = 120$ to $t = 220$ at (b)   $\Omega_{\mathrm{f}}=0.825$, (c)  $\Omega_{\mathrm{f}} = 0.85$, (d) at $\Omega_{\mathrm{f}}=0.9$. At smaller frequencies the plots show $k$ scaling at smaller $k$ and for larger frequencies show $k^{-3/2}$ and $k^{-7/2}$ scaling at larger $k$ values.}
\label{fig:latckinspectra}
\end{figure}
\begin{figure}[!ht]
\centering
\includegraphics[width=\linewidth]{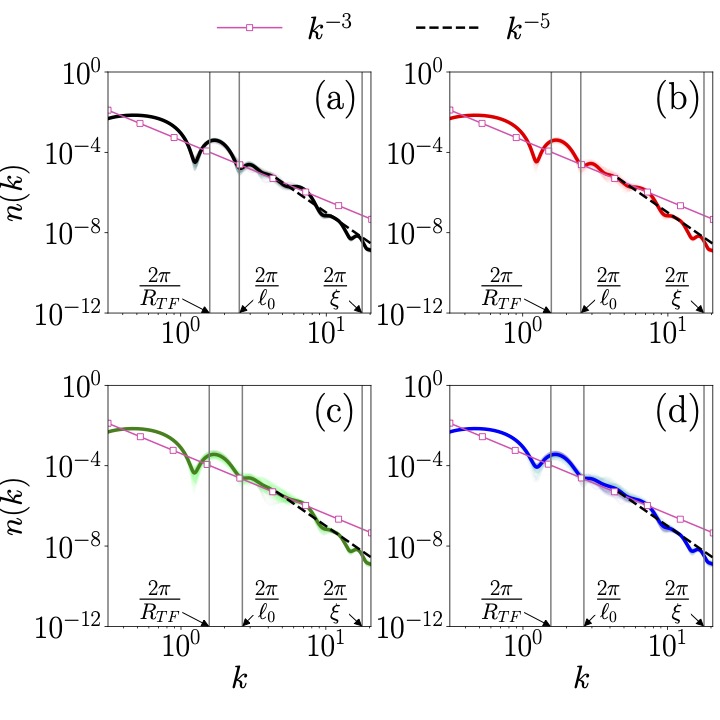}
\caption{Time averaged density spectra for the vortex-lattice case: (a) $t=280$ to $t=380$ at $\Omega_{\mathrm{f}}=0.8$, for   $t=120$ to $t=220$ at (b) $\Omega_{\mathrm{f}} = 0.825$, (c)  $\Omega_{\mathrm{f}} = 0.85$, (d)  $\Omega_{\mathrm{f}} = 0.9$.} 
\label{fig:latdenspectra}
\end{figure}

\begin{figure}[!ht]
\centering
\includegraphics[width=\linewidth]{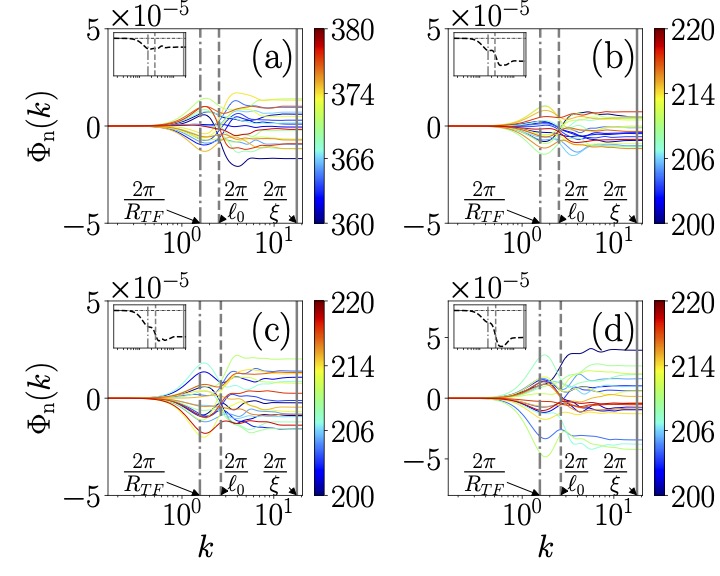}
\caption{Density flux profiles for the vortex-lattice case: (a) $t=360$ to $t=380$ at $\Omega_{\mathrm{f}}=0.8$, for $t=200$ to $t=220$ at (b) $\Omega_{\mathrm{f}} = 0.825$, (c) $\Omega_{\mathrm{f}} = 0.85$, (d)  $\Omega_{\mathrm{f}} = 0.9$. The inset displays the time average of the density flux (magnitude $\sim 10^{-6}$) for each rotational frequency plotted in the range $2\pi/R_{TF} < k < 2\pi/\xi$.
}
\label{fig:latdenflux}
\end{figure}

In Fig.~\ref{fig:latdensities}, we show the snapshots of the condensate at different instant of time as the condensate was finally set to the final rotation with frequency $\Omega_f=0.85$ at $t=120$. At $t=0$ [cf. Fig.~\ref{fig:latdensities}(a)], we find the presence of pinned vortices inhibiting them from reaching the condensate boundaries. The condensate gets distorted at $t=120$. Beyond $t=120$, the sustained rotation with frequency $\Omega_{\mathrm{f}} = 0.85$ introduces more vortices in the condensate, resulting in the enlargement of the condensate. During this time, very few vortices get clustered around the central barrier, while the rest of the vortices start spiralling around, thereby producing a disordered lattice [see Figs.~\ref{fig:latdensities}(c)-(d)]. 

Further, to get an insight into the transient states of the turbulence for the vortex-lattice initial state, we illustrate the time evolution of the mean angular momentum ($\langle L_z \rangle$) for different final angular frequencies ($\Omega_f=0.6$, $0.8$, $1.0$, and $1.2$) in Fig.~\ref{fig:amlattice}. Like vortex-less case for this initial state we also find that the $\langle L_z \rangle$ increases with an increase in $\Omega_{\mathrm{f}}$. The angular momentum attains to the quasi-static at shorter times, however, keep on gradually increasing due to the persistent rotation for $0.8 \lesssim \Omega_{\mathrm{f}} \lesssim 1$ . However, for $\Omega_{\mathrm{f}}=1$, $\langle L_z \rangle$ remains stable for a duration  $t \sim 200-1000$. Beyond that period $\langle L_z \rangle$ starts increasing profusely and starts having higher higher $\langle L_z \rangle$ than those for  $\Omega_{\mathrm{f}}=1.2$. This drastic change in angular momentum for the situation when the rotating frequency resonates with the trap frequency may be connected with the presence of large number of degenerate Landau levels at this frequency~\cite{Fetter2009}.

Figure \ref{fig:latikinspectra} illustrates the scaled incompressible kinetic energy spectra for the vortex lattice initial state at different rotational frequencies $\Omega_{\mathrm{f}} = 0.85, 0.825, 0.85$ and $0.9$ in the wavenumber space. The corresponding characteristics scales at different $\Omega_f$ have been provided in Table~\ref{tab:vlattice_scales}. Like the vortex less initial state for all the frequency $\Omega_{\mathrm{f}}$ the spectrum at large wave number follows $k^{-3}$ a typical behaviour for the energy spectrum due to the presence of the vortex core~\cite{Bradley2012}. For $k$ values $k > 2\pi/\ell_0$, where the dynamics of individual vortices dominate, the scaled spectrum fits well with $k^{-1}$ a typical feature of the Vinen-like turbulence and indicative of vortices clustering together without combining to form a larger vortex~\cite{Marino2021}. This $k^{-1}$ scaling is also shown to be accompanied by 2D weak wave turbulence confirmed by the compressible kinetic energy spectra (see Fig.~\ref{fig:latckinspectra}), which consistently show $k^{-3/2}$ and $k^{-7/2}$ scaling laws. The Vinen or ultraquantum scaling of $k^{-1}$ suggests no accumulation of kinetic energy at large scales and a direct cascade of energy in the region $2\pi/\ell_0 < k < 2\pi/\xi$ onward. 

Figure~\ref{fig:latkinflux} shows the flux of the incompressible component of the kinetic energy for the vortex-lattice initial state. Panels (a)-(e) represent the flux for the condensate as it is rotated with the rotation frequency $\Omega_{\mathrm{f}} = 0.85, 0.825, 0.85$ and $0.9$, respectively. The different colors represent the instants at which we plot the flux as indicated in the color bar. Like the vortex-less initial state, this case also exhibits oscillation in the flux profile with frequency $2.0\omega$ for all rotational frequencies of the condensate. However, the time-averaged flux shown as the thick dashed black line exhibits a constant positive flux in the UV range, indicating the forward cascade of incompressible kinetic energy.

The compressible spectra of this the vortex initial state as shown in Fig.~\ref{fig:latckinspectra} exhibits $k^1$ scaling at small wave number while exhibits  $k^{-3/2}$ and $k^{-7/2}$, respectively in the $k > 2\pi/\ell_0$ and UV ranges for all the rotation frequency $\Omega_{\mathrm{f}} = 0.85, 0.825, 0.85$ and $0.9$.

\begin{figure}[!ht]
\centering
\includegraphics[width=\linewidth]{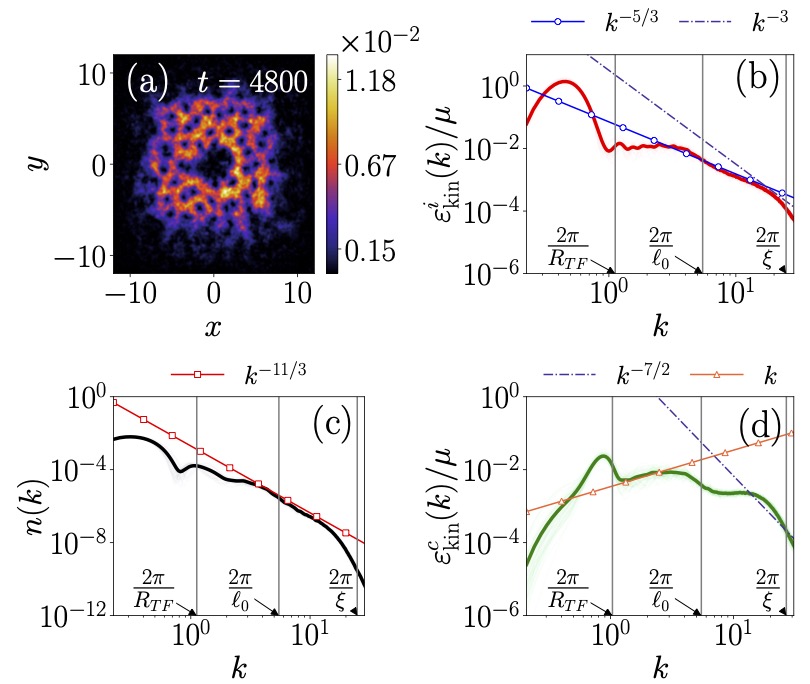}
\caption{Plots of (a) density and spectral profiles of (b) incompressible kinetic energy, (c) density, and (d) compressible kinetic energy averaged over the time interval $t=3600$ to $t=4800$ for vortex-lattice rotating at $\Omega_{\mathrm{f}}=1.0$ with $R_{TF}=6.14$, $\ell_0=1.15$, $\xi=0.23$. 
}
\label{fig:latspecden}
\end{figure}
\begin{figure}[!ht]
\centering
\includegraphics[width=\linewidth]{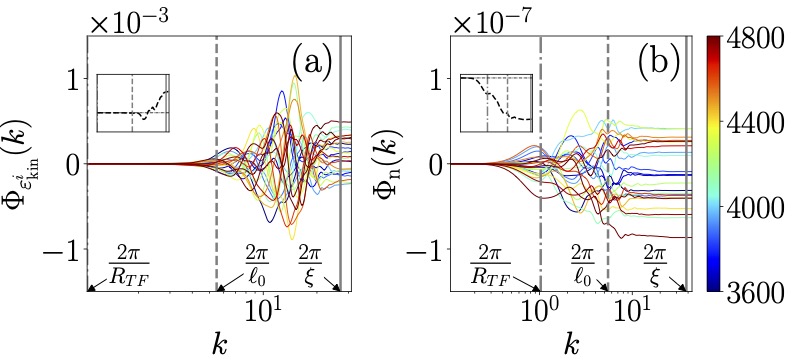}
\caption{Plots of (a) incompressible kinetic energy and (b) density fluxes for vortex-lattice rotating at $\Omega_{\mathrm{f}}=1.0$ for the time interval $t=3600$ to $t=4800$. For (a) Both the inset and the main plot are shown in the range $2\pi/R_{TF} < k < 2.3\pi/\xi$, while (b) the inset and main plot are in the range $0.2\pi/R_{TF} < k < 2.3\pi/\xi$}
\label{fig:latlongflux}
\end{figure}

In Fig.~\ref{fig:latdenspectra}, we plot the density spectra for the vortex-lattice initial state for the, which scales as $k^{-3}$ for $k^{-1}$ scaling in incompressible spectra. For all the final rotating frequencies ($0.8 < \Omega_{\mathrm{f}}<0.9$), which display scaling in the transient regime, it does not show any significant and sustained turbulence, even when the condensate energy settles. This is due to the perturbations not extending all the way to the condensate boundary and being unable to develop significant compressible turbulence. However, when rotated at sufficiently higher frequencies for longer time periods, we observe consistent and sustained turbulent behaviour. We notice strong quantum turbulence with a pronounced scaling law for the final rotating frequency of $\Omega_{\mathrm{f}} = 1.0$, as shown in Fig.~\ref{fig:latspecden}(b), where the $k^{-5/3}$ and $k^{-3}$ scales indicate the onset of the Kolmogorov cascade accompanied by an enstrophy cascade. We also note that the accompanying density spectra, as shown in Fig.~\ref{fig:latspecden}(c), display $k^{-11/3}$ scaling. However, the compressible spectra shows linear increasing behaviour with $k$ in the IR range while $k^{-7/2}$ in the UV range same as those observed with the vortex less cases [see Fig.~\ref{fig:latspecden}(d)]. Further we complement the chaotic observation of the condensate through the energy density representation in the real space in the Fig.~\ref{fig:latlongienerden}-\ref{fig:latlongcenerden}. Note that compared to the Kolmogorov scaling observed in the vortex-less initial wave profile, the scaling behaviour, in this case, appears at larger $k$ values. This is further confirmed by the kinetic energy and density flux profiles in Fig.~\ref{fig:latlongflux}, which display positive and negative average values, respectively, over the $k$ range, indicating a direct energy cascade.

To further confirm the influence of the perturbation barrier on generating the turbulence in the condensate, we have also considered the condensate along with the quasiperiodic central barrier as illustrated in Fig.~\ref{fig:barrier}(b). We find that the presence of a quasiperiodic barrier does not bring much change in the dynamics of the condensate with both vortex-less and vortex-lattice initial states. With the quasiperiodic perturbation, we do not find any signature of turbulent fluctuation in the condensate. Both incompressible and compressible components of the kinetic energy also do not exhibit any scaling laws for various final rotation frequencies. However, one interesting observation is made at higher rotation frequencies closer to $\Omega_{\mathrm{f}} = 1.0$. For this, we notice the presence of strong turbulence in the condensate, which is quite evident from the power law behaviour of the incompressible kinetic energy in the $k<2\pi/\ell_0$ range that exhibits Kolmogorov-like scaling ($\varepsilon_{\mathrm{kin}}^i(k)\sim k^{-5/3}$) for both the initial state which is similar to that those a periodically perturbed barrier.  
 
\section{Summary and Conclusions}
\label{sec:summary}

In this paper, we investigated the impact of the perturbed central barrier on rotating Bose-Einstein condensates confined in harmonic traps. We found that a strong perturbation can trigger the formation of an ordered vortex lattice from a vortex-free condensate, showing a direct energy cascade with Kolmogorov scaling, and, in contrast, the initial wave profile with a vortex-lattice gets distorted during transient stages and exhibits a non-Kolmogorov scaling law within the inertial range and displays no consistent turbulence. But at a longer duration under higher rotation frequencies, we observe turbulent behaviour following Kolmogorov scaling. 

For both cases, we have confined our studies to a range of initial and final rotation frequencies mainly based on the appearance of scaling laws and the onset of consistent turbulent dynamics. More generally, for rotating frequencies greater than the trap frequency, a condensate undergoing rotation will undergo constant expansion followed by an uncontrolled generation of vortices. This region presents with no interesting spectral profiles as the vortex core scaling of $k^{-3}$ is totally absent. Although the angular momentum expectation values provide an idea regarding the dynamical behaviour in both cases, we confirm and quantify its exact nature via spectral analysis, energy density and flux profiles of the kinetic energy components.  

When subjected to turbulent rotation, the vortex profile establishes an ordered and symmetric vortex lattice whose existence and scale depend on the final rotating frequency, $\Omega_{\mathrm{f}}$. With faster rotation, the condensate can sustain more vortices, which eventually become attracted toward the central barrier. This process, accompanied by strong compressible turbulence, is indicated by the scalings of $k$ in the compressible kinetic energy spectra and $k^{-11/3}$, and $k^{-5/3}$ at longer length scales ($k < 2\pi/\ell_0$) for density and incompressible spectra, respectively. To further confirm the nature of the energy cascade, we computed the energy fluxes which exhibit direct cascade.

In the case of condensate comprising of vortex lattice initial state (containing a sufficient number of vortices), the influence of the perturbed barrier on the lattice structure leads to distinct scaling behaviour. The presence of the vortex lattice is equivalent to initiating a condensate with energy injected at smaller scales. According to Barenghi~\cite{Barenghi2023}, this constitutes a recipe for generating a disordered vortex lattice, characterized by $\varepsilon_{\mathrm{kin}}^i(k) \sim k^{-1}$ scaling in the incompressible kinetic energy spectra. The $k^{-1}$ scaling in the incompressible spectra and $k^{-3}$ scaling in density spectra are indicative of weak wave turbulence, a fact reaffirmed by the presence of $k^{-3/2}$ and $k^{-7/2}$ scaling in the compressible spectra \cite{Marino2021}. These scaling tendencies are observed for final rotational frequencies within the range $0.8 < \Omega_{\mathrm{f}} < 0.9$.
In contrast to the scenario without vortices, choosing $\Omega_0$ value to be greater than the critical frequency $\Omega_{\mathrm{c}}$ of the condensate causes the condensate to expand and introduce vortices, reducing the perturbation amplitude relative to the condensate size. As a result, the perturbations are unable to generate sufficient compressible turbulence. Despite the appearance of $k^{-1}$ scaling in the and $k^{-3/2}$,$k^{-7/2}$ scaling behaviour in the incompressible and compressible spectra respectively in the transient time iterations, the lack of sufficient turbulence shows no consistent scaling behaviour at longer durations, where the condensate energy settles. By limiting the perturbation amplitude in such a manner, we only observe turbulence by rotating the condensate at higher frequencies for longer durations, where the condensate displays $k^{-5/3}$ Kolmogorov scaling.

In conclusion, our study highlights the behaviour of a perturbed barrier when subjected to rotation. In addition to playing a crucial role in generating turbulence, the perturbed barrier also significantly influences the nature of this turbulence, depending on the choice of initial wave profiles, especially in the presence of vortices. Under the influence of these perturbations combined with a centered axis of rotation, vortices tend to cluster around the barrier. Forcing vortices into the initial condensate causes the barrier to behave differently by distorting the vortex lattice rather than creating a settled vortex cluster. Our method provides a novel mechanism to generate turbulence and alter the behaviour of the condensate in a rotating frame with a barrier whose geometry and strength can be fine tuned. This study can be simulated and experimentally realized for a given BEC, provided the barrier has sufficient perturbation amplitude, barrier radius, and strength.

Finally, our present study is relevant for exploring the dynamical evolution of fluids inside boundary layers between the superfluid cores inside pulsars and the ambient normal matter, where the observed glitch phenomena are said to occur. Once the rotating superfluid core jumps promptly into the next lower energy state abruptly, a certain amount of rotational energy is ejected into the ambient medium, setting the medium in the boundary layer in a turbulent mode. Following Refs.~\cite{Hujeirat2018, Hujeirat2019}, such events are predicated to reoccur several billion times during a pulsar's lifetimes; hence, the core serves as a forcing term for generating turbulence, as alluded to in the present study.

\acknowledgments
A.S. acknowledges MoE RUSA 2.0 (Bharathidasan University - Physical Sciences) for financial support. The work of P.M. is supported by DST-SERB under Grant No. CRG/2019/004059 and MoE RUSA 2.0 (Bharathidasan University - Physical Sciences).

\section*{Author Declarations}

\subsection*{Conflict of Interest}
\noindent The authors have no conflicts to disclose.

\subsection*{Author Contributions}
\noindent\textbf{Anirudh Sivakular:} 
Conceptualization (equal); Investigation (equal); Validation (equal); Visualization (equal); Writing – original draft (equal).
\textbf{Pankaj Kumar Mishra:}  Conceptualization (equal);  Investigation (equal); Supervision (lead); Writing - original draft (equal): Writing – review \& editing (equal).
\textbf{Ahmad A. Hujeirat:}  Conceptualization (equal);  Writing – review \& editing (equal).
\textbf{Paulsamy Muruganandam:}  Conceptualization (equal);  Investigation (equal); Supervision (lead); Writing - original draft (equal): Writing – review \& editing (equal).

\section*{Data Availability Statement}
The data supporting the findings of this work can be generated by solving the equations with the methods provided. Additionally, the data can be obtained from the corresponding author upon reasonable request.

\appendix
\counterwithin{figure}{section}
\section{Incompressible and compressible energy densities}
\label{app1}

In addition to the spectral profiles discussed in subsections \ref{sec:vlesscase} and \ref{sec:vlatticecase}, it is worth examining the incompressible and compressible density profiles of kinetic energies in real space. In this appendix, we illustrate the incompressible and compressible density profiles of kinetic energies for both the vortex-less and vortex lattice cases.

\begin{figure}[!ht]
\centering
\includegraphics[width=\linewidth]{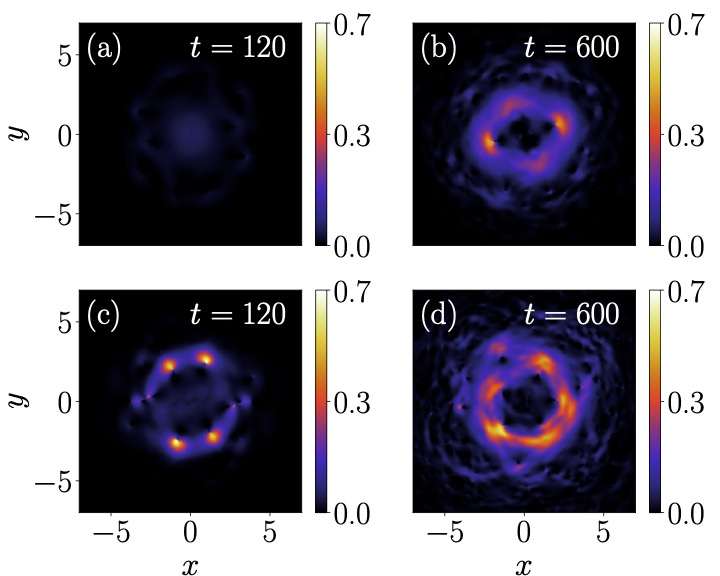}
\caption{ 
Snapshots of the incompressible energy density for the vortex-less case: $\Omega_{\mathrm{f}} = 0.85$ at (a) $t=120$ and (b) $t=600$, and $\Omega_{\mathrm{f}} = 1.0$ at (c) $t=120 $ and (d) $t=600$.
} 
\label{fig:ienerden}
\end{figure}

In Fig.~\ref{fig:ienerden}, we show the snapshot of the density profile of the incompressible kinetic energy in real space at initial ($t=120$) and final ($t=600$) instant of time for $\Omega_{\mathrm{f}} = 0.85$ and $\Omega_{\mathrm{f}} = 1$.  We can observe that the incompressible energy, initially localized at the barrier, is relatively small [See Figs.~\ref{fig:ienerden}(a), (c)].  

As turbulence sets in, the energy becomes concentrated around the edges of the barrier, indicating the presence of vortices. For higher rotation frequencies [see Figs.~\ref{fig:ienerden}(c)-(d)], the density distribution follows a similar trend, but at later times, there is a band of high-energy density, indicating the presence of several vortices clustered around the barrier.
\begin{figure}[!ht]
\centering
\includegraphics[width=\linewidth]{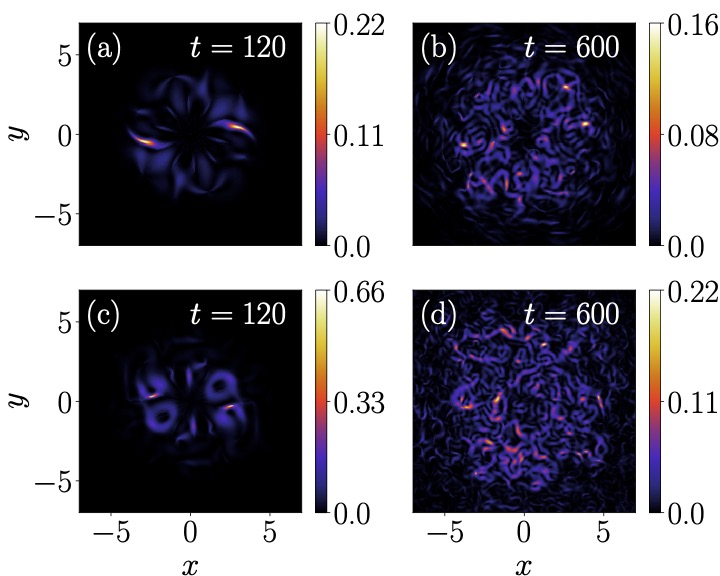}
\caption{Snapshots of the compressible energy density for the vortex-less case: $\Omega_{\mathrm{f}} = 0.85$ at (a) $t=120$ and (b) $t=600$, and $\Omega_{\mathrm{f}} = 1.0$ at (c) $t=120$ and (d) $t=600$. 
}
\label{fig:cenerden}
\end{figure}%
The initial time snapshots illustrate the localization of compressible energy near the barrier edge. This emphasizes the role of the barrier in inducing turbulence as it compresses the condensate between the perturbations. %
As time progresses, the compressible turbulence becomes evenly distributed within the condensate surrounding the barrier [See Fig.~\ref{fig:cenerden}(b) and (d)]. This observation shows significant phonon activity and thermalization, also confirmed by the compressible energy spectra.

This observation confirms what we have noticed in the spectral profiles and their time derivatives as the energy becomes more concentrated toward the healing length or vortex-size scales. In Fig.~\ref{fig:cenerden}, we show the snapshots of the compressible kinetic energy densities for $\Omega_{\mathrm{f}} = 0.85$ (upper panel) and $\Omega_{\mathrm{f}} = 1.0$ (lower panel) at time $t=120$ and $t=600$. %
\begin{figure}[!ht]
\centering
\includegraphics[width=\linewidth]{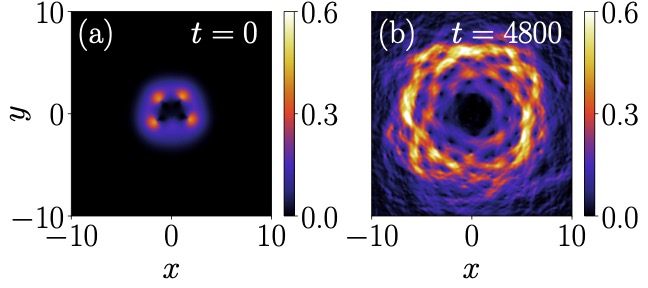}
\caption{Snapshots of the incompressible energy density for the vortex-lattice case: $\Omega_{\mathrm{f}} = 1.0$ at (a) $t=0 $ and (b) $t=4800 $ }
\label{fig:latlongienerden}
\end{figure}%
\begin{figure}[!ht]
\centering
\includegraphics[width=\linewidth]{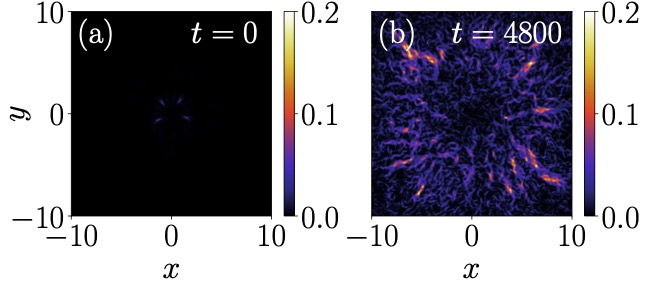}
\caption{Snapshots of the compressible energy density for the vortex-lattice case: $\Omega_{\mathrm{f}} = 1.0$ at (a) $t=0 $ and (b) $t=4800 $ }
\label{fig:latlongcenerden}
\end{figure}%
The density profiles of the kinetic energy components further confirm the disordered nature of the turbulence. The incompressible kinetic energy densities shown in Fig.~\ref{fig:latlongienerden}, for $\Omega_{\mathrm{f}} = 1.0$ show an ordered vortex lattice at the initial stage, which, when rotated at such a high frequency develops considerable turbulence with the incompressible density being similar to that of the vortex-less case but larger and more pronounced. The compressible density profiles shown in Fig.~\ref{fig:latlongcenerden} for the same parameters show no significant development of compressible turbulence at initial stages but later distribute itself more evenly at the highly turbulent state.

%

\end{document}